# DSC-MRI derived relative CBV maps synthesized from IVIM-MRI data: Application in glioma IDH mutation status identification

Lu Wang[1], Zhen Xing[2], Congbo Cai[1], Zhong Chen[1], Dairong Cao[2*], Shuhui Cai[1*]

[1]Department of Electronic Science, Fujian Provincial Key Laboratory of Plasma and Magnetic Resonance Xiamen University, Xiamen, Fujian, China

[2]Department of Radiology, First Affiliated Hospital of Fujian Medical University, Fuzhou, Fujian, China

## Introduction

Dynamic susceptibility contrast magnetic resonance imaging (DSC-MRI) involves the injection of gadolinium-based contrast agent (GBCA) to dynamically alter the $T_2$ or $T_2^*$ transverse relaxation rate of tissue for quantitative perfusion information acquisition. DSC-MRI-derived cerebral blood volume (CBV) has been proven of ignorable significance in providing information about tumor malignancy and has been applied to brain tumor treatment monitoring [1-3]. Since different DSC-MRI postprocessing algorithms result in different CBV values, relative CBV (rCBV) is more commonly used [4]. Isocitrate dehydrogenase (IDH) mutation status is a key prognostic factor for glioma, which has been discovered to be associated with a distinct angiogenesis transcriptome signature that is noninvasively predictable with rCBV imaging [5-7]. A vascular habitat analysis based on DSC-MRI has concluded that the rCBV could discriminate IDH mutation from wildtype [8]. In all, DSC-MRI is of considerable significance in clinic, especially for the rCBV parameter representing neovascularization. However, although widely acknowledged as safe, GBCA is still associated with potential limitations. First, the GBCA injection increases the cost of brain MRI examination with the requirement of physician monitoring. Second, pregnant patients are not suggested to accept MR examinations with GBCA administration. Third, patients with impaired kidney function suffer from impaired clearance of gadolinium, leading to tissue accumulation of dissociated gadolinium, whose clinical significance remains unknown [9-11]. Fourth, the GBCA injection is not recommended for treatment evaluation that requires follow-up. Hence, finding alternative ways to obtain DSC-MRI derived rCBV without GBCA injection is vital.

Theoretically, CBV could be represented as the product of intravoxel incoherent motion magnetic resonance imaging (IVIM-MRI) derived perfusion fraction $f$ (IVIM-$f$) and tissue NMR-visible water content fraction [12]. IVIM-MRI is a non-invasive perfusion imaging technique from which $f$ could be obtained by fitting a bi-exponential model using multi diffusion weighting ($b$ value) diffusion weighted imaging (DWI) data [13]. Although IVIM-$f$ was identified to correlate well with DSC-MRI-derived rCBV [13; 14], to the best of our knowledge, no existing work achieves high-quality conversion from IVIM-MRI to DSC-MRI-derived rCBV. One important reason may be that high-quality IVIM-$f$ is challenging to obtain, especially for in vivo brains with lower IVIM-$f$ [15; 16].

We assumed that DSC-MRI-derived rCBV maps could be converted from IVIM-MRI data. Medical image modality conversion, which is similar to optical image style transfer [17], has already been explored in MRI and computed tomography (CT) interconversion [18] and conventional MRI modality interconversion [19; 20] with generative adversarial networks (GANs). Based on GANs, synthetic FLAIR images could be generated from DWI, resulting in reduced imaging protocol time, which is of much value for acute ischemic stroke patients [21]. Apart from GANs, the pix2pix network is also an alternative with which synthetic rCBV could be generated from dynamic contrast-enhanced MRI (DCE-MRI), enabling robust brain-tumor perfusion imaging of DSC and DCE parameters with a single GBCA administration [22]. Furthermore, the DSC-MRI-derived perfusion parameters have also been successfully generated from contrast-enhanced MR angiography (MRA) with high pixel-wise and structure similarity using U-Net and its variants [4]. U-Net has also been applied to synthesize high-quality $T_1$-weighted ($T_1$w) images directly from diffusion data and help improves the brain segmentation accuracy in diffusion MRI data analysis [23]. Besides, Calabrese et al. and Chung et al. have validated the feasibility of simulated contrast-enhanced $T_1$w brain and breast MRI generated from multiple pre-contrast sequences with a three-dimensional fully convolutional deep neural network [24; 25]. Inspired by the mentioned works, we employed deep neural networks to verify our assumption.

The rCBV synthesis was ever fulfilled with DCE-MRI or MRA data as neural network input [4; 22]. Essentially, these works mainly focus on obtaining DSC-MRI-derived parameters from post-contrast dynamic $T_1$w scans. Therefore, the injection of GBCA still cannot be avoided. We aim to propose an rCBV synthetic method for which GBCA is out of demand to benefit kidney-impaired and pregnant patients and treatment follow-up evaluation. We first established and verified the proposed framework on IDH wildtype glioma patients. To evaluate the generalizability, we tested the model trained on IDH wildtype glioma data with IDH mutation glioma data and did structure similarity and consistency analysis with DSC-MRI derived real rCBVs. Furthermore, since IDH mutation gliomas have been proven to show lower CBV than IDH wildtype gliomas [26], we used the synthetic rCBV for glioma IDH mutation status identification of 40 patients, further validating the potential clinical value of our proposed method.

## Materials and methods

This study was approved by the Branch for Medical Research and Clinical Technology Applications, the Ethics Committee of First Affiliated Hospital of Fujian Medical University, and informed consent was obtained.

### Patient inclusion

From September 2016 to December 2018, 146 patients with pathologically confirmed glioma underwent multimodality MR examinations, including DSC- and IVIM-MRI. Exclusion criteria were any severe distortion of DWI images or failed rCBV calculation, including abnormally high signal intensity in certain brain regions and signal over-attenuation in tumor regions and big vessels. Detailed patient inclusion diagram is shown in Figure 1.

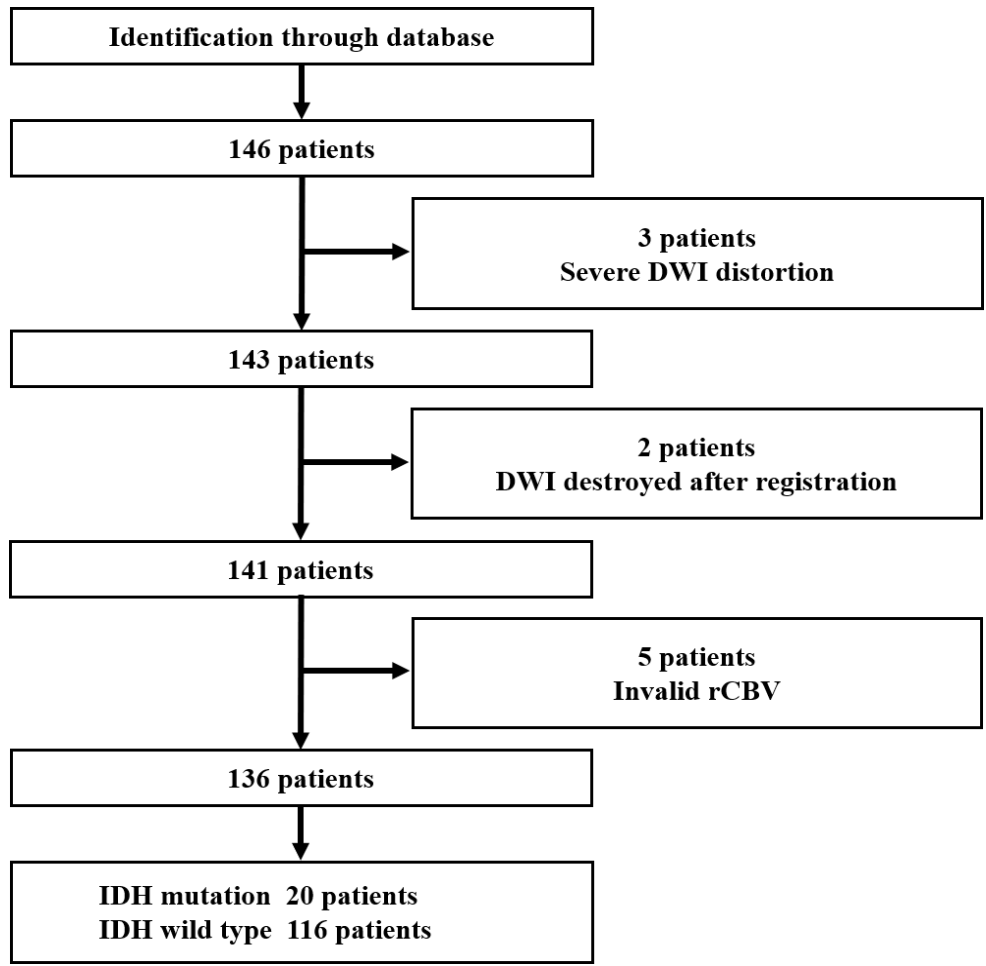

Figure 1. Flowchart for patient inclusion. DWI = diffusion-weighted imaging, rCBV = relative cerebral blood volume.

## MR imaging

Both DSC- and IVIM-MRI data were acquired on 3T SIEMENS Skyra scanners with 16-channel head-neck coils. Detailed imaging protocol is as follows: (1) IVIM-MRI: Spin-echo EPI sequence, 220 × 220 mm$^2$ field of view, 5 mm slice sickness, 6 mm spacing between slices, TE/TR = 68/4100 ms, echo train length = 54 ms; voxel size was 0.73 × 0.73 × 5.0 mm$^3$, $b$ values were 0, 50, 100, 150, 200, 300, 400, 600, 800, 1000 s/mm$^2$ and were repeated 2, 2, 2, 2, 2, 2, 2, 2, 3, 3 times; (2) DSC-MRI: Gradient-echo EPI sequence, 220 × 220 mm$^2$ field of view, 5 mm slice sickness, 6 mm spacing between slices, TE/TR = 30/1600 ms, echo train length = 63 ms; voxel size was 1.7 × 1.7 × 5.0 mm$^3$. The images during the first three phases were obtained before the GBCA was injected to establish a non-enhanced baseline. When the scan reached the fourth phase of DSC-MRI, a bolus of gadobenate dimeglumine was injected intravenously at a dose of 0.1 mmol/kg and a rate of 5 ml/s, followed by 20 ml of saline injected at the same rate. A total of 1,200 images were obtained in 96 seconds (20 sections, 60 phases). The real rCBV maps were established by applying a single-compartment model and an automated arterial input function. (3) T$_2$w: Spin-echo sequence, 220 × 220 mm$^2$ field of view, 5 mm slice sickness, 6 mm spacing between slices, TE/TR = 125/6000 ms, echo train length = 54 ms; voxel size was 0.57 × 0.57 × 5.0 mm$^3$.

**Proposed scheme**

The following formula can illustrate the relationship between CBV and IVIM-*f* [12]:

$$f = V_c / V_W = V_c / f_W V = CBV / f_W \tag{1}$$

knowing the tissue NMR-visible water content fraction ($f_W = V_W/V$), $f$ (IVIM-*f*) represents the ratio of the NMR-visible water volume in the capillary compartment ($V_c$) to the total NMR-visible water voxel volume $V_W$, and can be converted into milliliters of capillary blood per 100 g of tissue (*CBV*). The IVIM biexponential model is as follows:

$$S_{IVIM}(b) = S_0[(1-f)e^{-bD} + fe^{-bD^*}] \tag{2}$$

where $S_0$ and $S_{IVIM}(b)$ are the signals obtained without and with diffusion weighting, $f$ is the perfusion fraction, $D$ is the diffusion coefficient and $D^*$ represents the pseudo-diffusion coefficient.

Figure 2 shows the proposed framework for rCBV synthesis. Multi-*b*-value DWI images (providing the information of *f* in Eq. 1) and T$_2$-weighted (T$_2$w) images were employed as the neural network input, together with rCBV maps obtained from the software syngo MR E11 embedded in 3T SIEMENS Skyra scanners as the training labels (real rCBV hereafter). All Real rCBV maps and multi-*b*-value DWI images were automatically registered to T$_2$w images of each patient with MATLAB toolbox Statistical Parametric Mapping 12 (SPM12). It is worth noticing that each slice of the T$_2$w images and real rCBV maps was normalized by setting the maximum value of each image as the baseline, and the IVIM-MRI data were normalized by setting the maximum value of *b* = 0 image as the baseline to get a clear lesion appearance for high *b*-value DWI images. T$_2$w images were used for obtaining water content fraction information $f_W$.

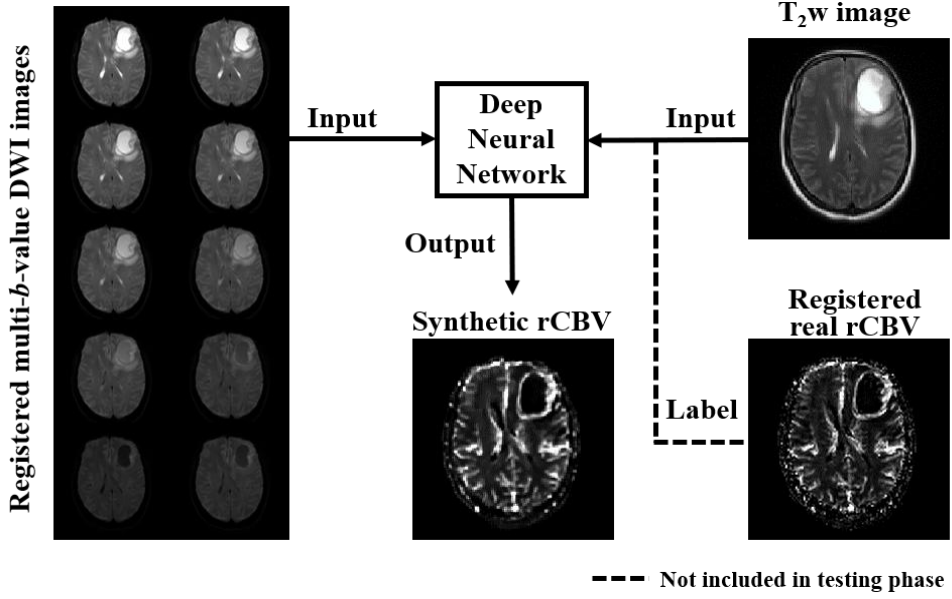

Figure 2. The framework of the generation of DSC-MRI derived rCBV maps from IVIM-MRI data. Five-level U-Net with perception loss is employed for nonlinear mapping. $S_0$ images mean $b = 0$ DW images. The dashed line represents the registered real rCBV maps included in the training samples while absent in the testing samples.

## Deep neural network training

A 5-level U-Net was used to fulfill the nonlinear mapping with the following loss function:

$$L1Loss(GT, pred) = mean(|GT - pred|) \tag{3}$$

$$L1Loss_{perceptual}(GT, pred) = L1Loss(perceptual(GT), perceptual(pred)) \tag{4}$$

$$Loss_{network} = L1Loss(GT, pred) + 0.1 \times L1Loss_{perceptual}(GT, pred) \tag{5}$$

where $GT$ is the ground truth (real rCBV), and $pred$ represents the network prediction (synthetic rCBV). $L1Loss$ was calculated as the mean absolute error (MAE). The $L1Loss_{perceptual}$ was calculated as the MAE of the extracted feature maps of real and synthetic rCBV maps from a pre-trained VGG19 network, where the feature maps represented important structures and shapes of the rCBV maps. The $L1Loss$ and $L1Loss_{perceptual}$ were used to maximize the pixel-wise and structure-wise similarity between neural network prediction and ground truth, respectively. Data augmentation, including image rotation, shrink and amplification, was done to obtain a larger quantity of training samples. Patient data were randomly divided into training, validation, and testing sets; 3000 slices were available for training and 300 for validation in the final. It is noteworthy that limited by case number, all the data for training and validation (96 patients) were IDH wildtype, the remaining IDH wildtype (20 patients) and all IDH mutation data (20 patients) were for testing to test the generalization ability of the proposed framework. A batch size of 8 was adopted to avoid excessive memory burden during training. The iterations were set to 2000 epochs and the patch size was 96 × 96. We used the Adam optimizer with momentum parameters $β_1 = 0.9$ and $β_2 = 0.999$ to update neural network parameters. The learning rate was initially set to $10^{-4}$ and decreased by 20% after every 40000 iterations until the neural network converged. The training time was around 7 hours based on the PyTorch development platform using an NVIDIA GeForce RTX 2080Ti GPU.

**Image analysis**

The CBV measurement shows high variability due to physiological differences in patients, including cardiac output and hematocrit values. Thus, CBV is commonly calculated by normalizing to a reference tissue [27]. Contralateral centrum semiovale has been proven to provide the lowest observer variability as the reference tissue, but it is most visible in only one or two axial slices. So our region of interest (ROI) analysis follows Sanders et al. [22]. Tumor and contralateral normal-appearing white matter (NAWM) ROIs were manually drawn in the real rCBV maps of the 40 test patients by a neuroradiologist with four years of experience. There were three tumor ROIs and three NAWM ROIs drawn for each patient. Each ROI was approximately 0.25 cm$^2$ in size. Tumor ROIs were determined on hot spots in the real rCBV maps when available, or areas of clinical relevance based on anatomical images for each patient. Following Wu et al. [8], the postcontrast $T_1$-weighted and FLAIR images were used to guild ROI drawing to avoid areas of necrosis, cysts, or non-tumor macro-vessels. The ROIs were then copied to the corresponding synthetic rCBV and IVIM-*f* maps. For all three quantitative maps, mean values of three ROIs were averaged for tumor and white matter (WM), respectively, to compute the tumor-to-white matter (T/WM) ratio. The T/WM ratio was used for consistency analysis and receiver operating characteristic (ROC) curve analysis of IDH mutation status differentiation. The lesion analysis was done on slices with the maximum tumor size determined on axial post-contrast $T_1$w images for each patient.

**Statistical analysis**

Statistical analysis was conducted with commercial software GraphPad Prism (Version 8.0a; GraphPad Software Inc) and MedCalc (Version 20.218; MedCalc Software). The structure similarity index measure (SSIM) between the real and synthetic rCBV was employed to determine the optimal neural network input combination. To quantitatively analyze the consistency between synthetic and real rCBV maps, linear regression analysis, Pearson correlation coefficient, and Bland-Altman analysis were done for T/WM ratios of real and synthetic rCBV maps. The difference between T/WM ratios of IDH mutation and IDH wildtype cases was evaluated with the unpaired *t*-test. ROC curves of glioma IDH mutation status identification were drawn, and areas under the ROC curves (AUCs) were computed and compared using the DeLong test. Specificity and sensitivity for glioma IDH mutation status identification were calculated at the cutoff 4.176 and 3.419 for real and synthetic maps, respectively. Statistical significance was indicated with a *P* value of 0.05.

## Results

### Synthetic rCBV under different input combinations

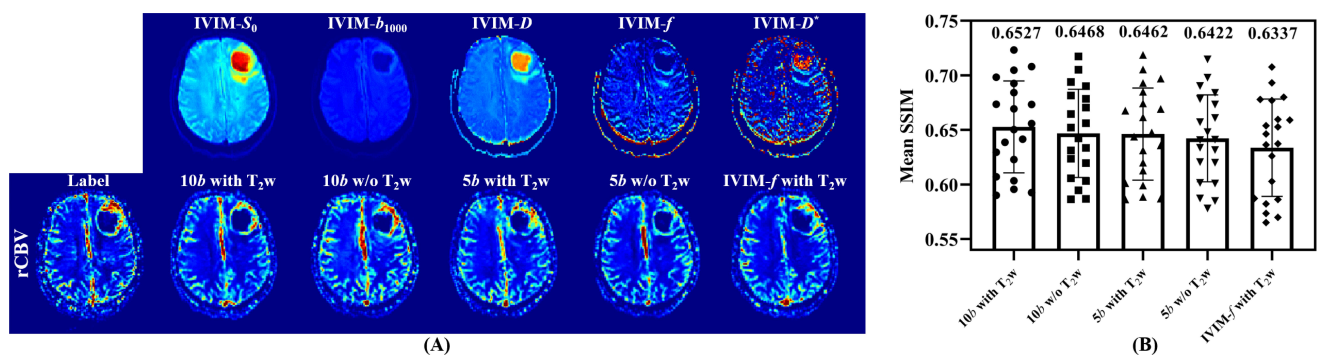

Figure 3. (A) First row: IVIM-MRI images (IVIM-$S_0$: $b = 0$ DWI; IVIM-$b_{1000}$: $b = 1000$ s/mm$^2$ DWI) and IVIM-MRI derived parameters (IVIM-$D$, IVIM-$f$, IVIM-$D^*$) obtained by fitting the biexponential model with 10-*b*-value DWI images; Second row: Synthesized rCBV maps with different input combinations (10*b*: $b = 0$, 50, 100, 150, 200, 300, 400, 600, 800, 1000 s/mm$^2$; 5*b*: $b = 0$, 50, 100, 150, 200 s/mm$^2$) of a specific slice for a patient with IDH wildtype glioma. (B) Mean SSIM of all slices preserved after data preprocessing and invalid data exclusion for 20 patients with IDH wildtype glioma.

Figure 3 illustrates the synthesized rCBV maps and calculated SSIM values with different input combinations. We determined the final input combination choice based on visual and structural similarities. From the perspective of visual similarity, as shown in Figure 3A, the prediction employing 10-*b*-value DWI and T$_2$w images as input, offers the most similar appearance to the realistic one. Compared to the predictions without T$_2$w images as input, the predictions utilizing T$_2$w images obtain more similar lesion contours to the realistic ones. From the perspective of structure similarity shown in Figure 3B, where each scatter represents the mean SSIM of all preserved slices for a single patient diagnosed with IDH wildtype glioma, it can be concluded that with the inclusion of T$_2$w images and high $b$-value DWI images, the structural similarity between synthetic and real rCBV maps increases in almost all IDH wildtype patients. It is worth noticing that with only IVIM-$f$ and T$_2$w images as input, the visual and structure similarities of the predictions are the worst. This is surprising, as the relationship between IVIM-$f$ and rCVB is theoretically more straightforward.

**Performance and consistency analysis of the proposed framework**

Figure 4 shows the synthetic rCBV maps of eight slices from healthy brain regions (Figure 4A) and eight slices from brains with pathologically confirmed IDH wildtype glioma (Figure 4B). For comparison, their corresponding real rCBV and IVIM-$f$ maps are provided. The synthetic and real rCBV maps are nearly indistinguishable, with similar high blood volume regions in slices without lesion and an apparent elevation of tumor rCBV compared to the surrounding parenchyma in glioma slices. Moreover, background noise is significantly reduced, and artifacts on phase encoding direction are thoroughly eliminated (Figure 4B, the eighth column) on synthetic rCBV maps. It is worth mentioning that the synthetic rCBV maps differ significantly from the IVIM-$f$ maps, with different signal intensity elevation of lesion areas as well as different local signal enhancement of normal tissue. This may explain why rCBV synthesized directly from IVIM-$f$ does not work well.

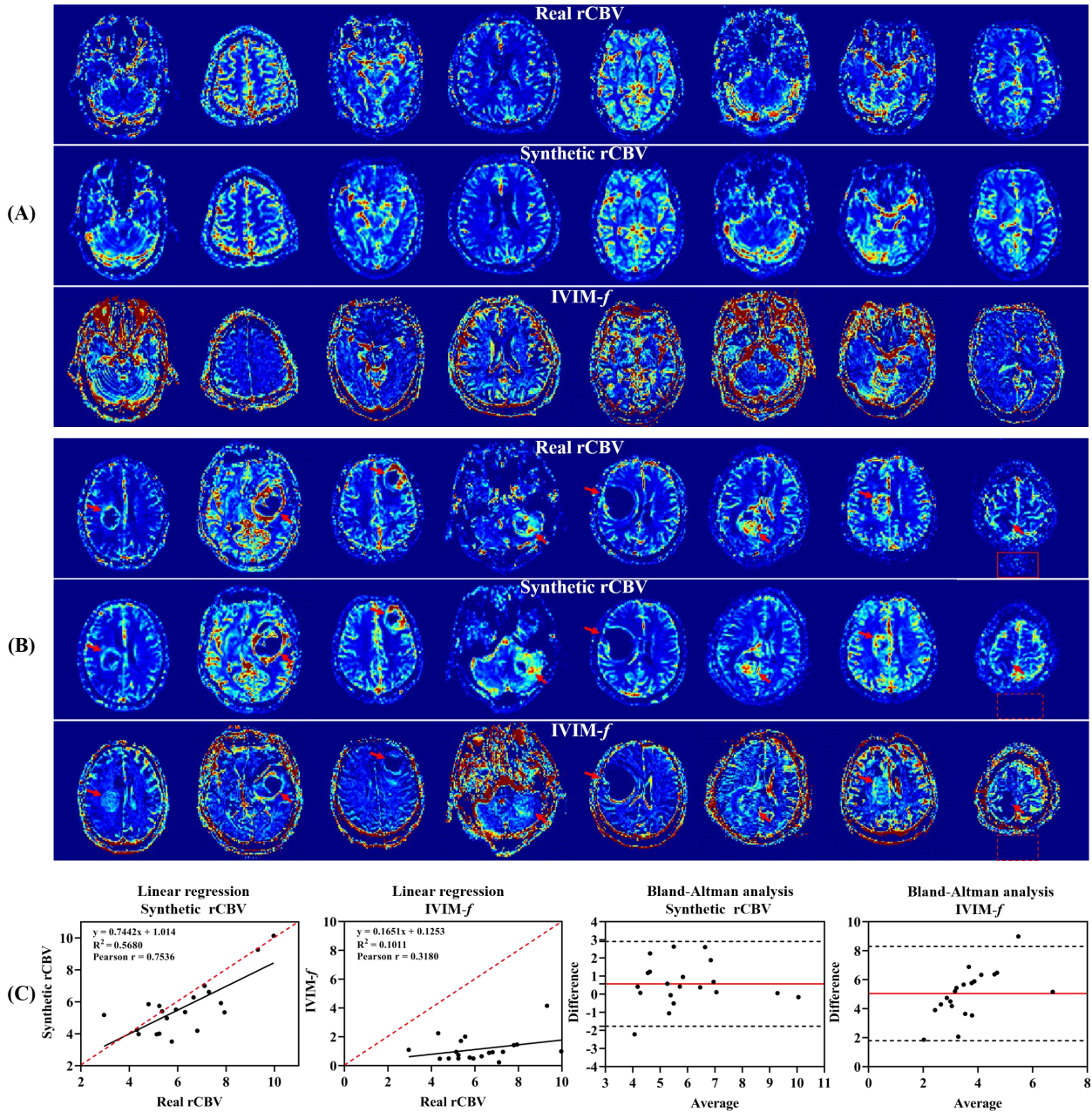

Figure 4. (A, B) The rCBV and IVIM-*f* maps of eight normal brain slices (A) and eight brain slices with pathologically confirmed IDH wildtype glioma (B). (C) Linear regression and Bland-Altman analysis plot between T/WM ratios of real rCBV and synthetic rCBV (or IVIM-*f*) for 20 patients diagnosed with IDH wildtype glioma. The red solid line represents the mean difference, and the upper and lower dash lines represent the 95% LoA. (Real rCBV: DSC-MRI derived rCBV maps obtained from the software syngo MR E11; Synthetic rCBV: IVIM-MRI derived rCBV maps obtained from the proposed rCBV generation method; IVIM-*f*: IVIM-MRI derived *f* maps.)

Results of the consistency analysis of T/WM ratios between real rCBV and synthetic rCBV of 20 IDH wildtype patients are shown in Figure 4C. It can be concluded that there is a linear relationship between them and could be represented as $(rCBV_{T/WM})_{synthetic} = 0.7442 \times (rCBV_{T/WM})_{real} + 0.04605$ ($R^2 = 0.5680$). And the Pearson correlation coefficient is 0.7536 ($P = 0.0001$). By comparison, the T/WM ratio of IVIM-*f* and real rCBV is not strongly related, with a Pearson correlation coefficient of 0.3180 ($P = 0.1718$). The Bland-Altman analysis of real and synthetic rCBV shows that almost all the data (19 of 20 [95%]) distribute within the 95% limits of agreement (LoA) with a mean difference of 0.5698, which is clinically acceptable [24]. For IVIM-*f*, although the Bland-Altman analysis shows that 95% data distribute within the 95% LoA, the mean difference (5.046) is significantly larger than synthetic rCBV. Real and synthetic rCBV maps of the whole brain for a patient diagnosed with IDH wildtype glioma could be found in Suppl. Figure S1.

**Generalizability of the proposed framework**

The synthetic rCBV maps are obtained from the neural network model trained with only IDH wildtype glioma data, and their visual qualities are similar to IDH wildtype predictions as mentioned. It is curcial whether the network can still work efficiently for IDH mutation glioma data. Figure 5A shows the real rCBV, synthetic rCBV and IVIM-*f* maps of two patients with pathologically confirmed IDH mutation glioma. Different from the IDH wildtype glioma displayed in Figure 4B, the IDH mutation gliomas shown in Figure 5A are almost invisible in the IVIM-*f* map but appear at the same location with matched elevated signal intensity in the synthetic rCBV map as the real rCBV map. Figure 5B shows that the SSIM analysis of the IDH mutation group is similar to the IDH wildtype group but with a certain lower SSIM. This may be due to network performance degradation caused by domain transferring (from IDH wildtype to IDH mutation).

Figure 5C illustrates the results of the linear regression analysis of T/WM ratios between real rCBV and synthetic rCBV of 20 IDH mutation patients. There is a linear relationship between them and it could be represented as $(rCBV_{T/WM})_{synthetic} = 0.7715 \times (rCBV_{T/WM})_{real} + 0.7143$ ($R^2 = 0.7980$). And the Pearson correlation coefficient is 0.8933 ($P < 0.0001$). By comparison, the T/WM ratio of IVIM-*f* and real rCBV is not strongly related, with a Pearson correlation coefficient of 0.1715 ($P = 0.4697$). The Bland-Altman analysis of real and synthetic rCBV in Figure 5D shows that almost all the data (19 of 20 [95%]) distribute within the 95% LoA with a mean difference of 0.2949, which is clinically acceptable. The Bland-Altman analysis results of IVIM-*f* in the IDH mutation group are similar to the IDH wildtype group. Real and synthetic rCBV maps of the whole brain for a patient diagnosed with IDH mutation glioma could be found in Suppl. Figure S2.

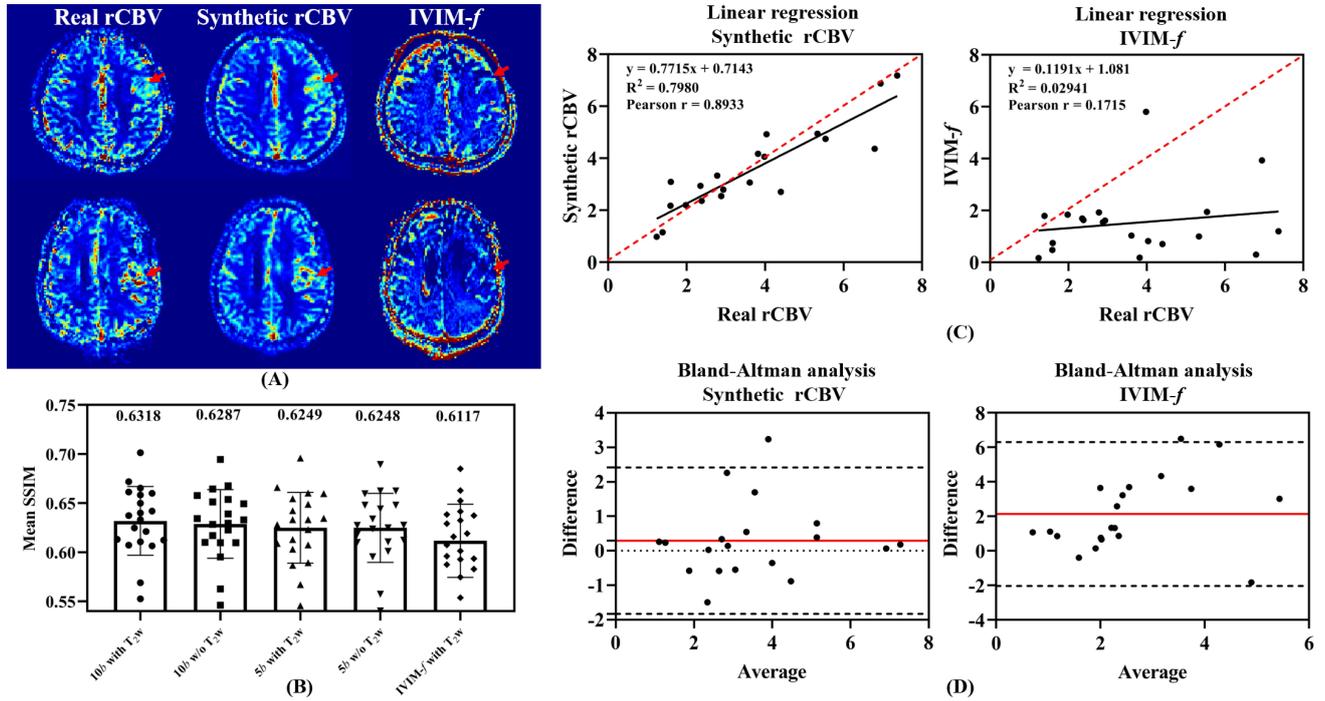

Figure 5. (A) The rCBV and IVIM-*f* maps of two patients diagnosed with pathologically confirmed IDH mutation glioma. (B) Mean SSIM of all slices preserved after data preprocessing and invalid data exclusion for 20 patients with IDH mutation glioma. (C) Linear relationship between T/WM ratios of real rCBV and synthetic rCBV (or IVIM-*f*) for 20 patients diagnosed with IDH mutation glioma. (D) Bland-Altman plot of T/WM ratios of real and synthetic rCBV (IVIM-*f*) for the same patients. The red line represents the mean difference, and the upper and lower dash lines represent the 95% LoA.

### Identification of glioma IDH mutation status

As shown in Figure 6A, the T/WM ratio of IDH mutation glioma is significantly lower than IDH wildtype glioma in both real ($P < 0.0001$) and synthetic rCBV ($P < 0.0005$) maps. Significant difference between the two glioma groups is not found in parameter IVIM-*f* ($P = 0.3102$). The ROCs for identification of IDH mutation and wildtype glioma in Figure 6B are not significantly different between real and synthetic rCBV maps (AUC = 0.8375 [95%CI: 0.687, 0.935] vs. AUC = 0.8325 [95%CI: 0.681, 0.932], respectively; $P = 0.9075$). Significant difference is found between the ROCs of real rCBV and IVIM-*f* maps, the IDH mutation status differentiation ability of IVIM-*f* is far weaker (AUC = 0.8375 [95%CI: 0.687, 0.935] vs. AUC = 0.59 [95%CI: 0.423, 0.743], respectively; $P = 0.0368$). Specificity and sensitivity for glioma IDH mutation status identification are 70% and 95% for real maps, 60% and 100% for synthetic maps.

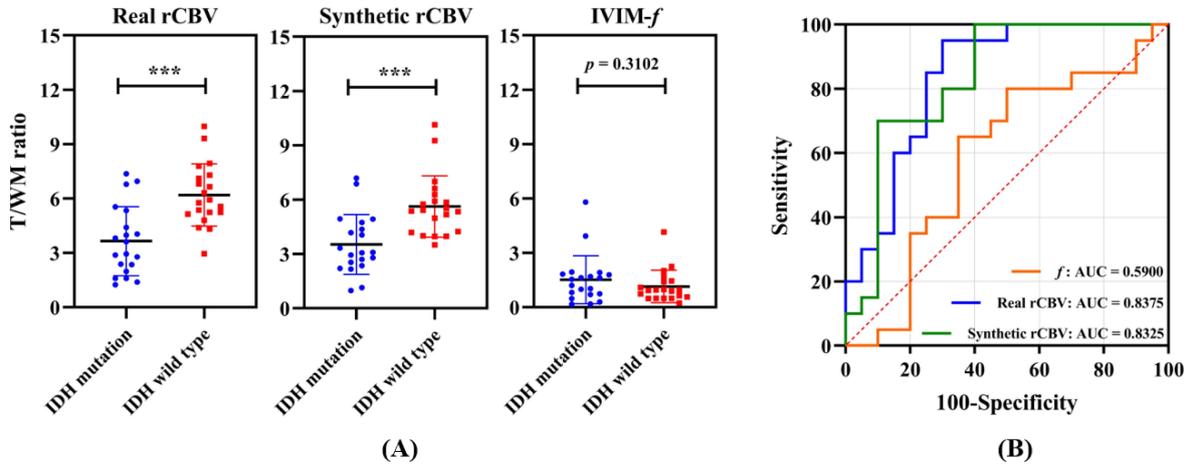

Figure 6. (A) T/WM ratios in IDH mutation and IDH wildtype glioma of IVIM-$f$, real and synthetic rCBV maps (unpaired $t$-test). (B) ROCs of T/WM ratios of IVIM-$f$, real and synthetic rCBV maps for glioma IDH mutation status identification.

## Discussion

This study proposes a method for generating DSC-MRI-derived rCBV maps from IVIM-MRI data based on a deep neural network. Consistency analysis and IDH mutation status identification ability analysis demonstrate the feasibility of synthesizing rCBV maps from IVIM-MRI. Significantly, the generalizability of the proposed framework is also verified with successful rCBV prediction on IDH mutation data and considerable IDH mutation status identification ability, using a deep neural network model trained purely on IDH wildtype data.

$T_2w$ images help to reveal lesion contour of synthetic rCBV (Figure 3A), improving the SSIM of synthetic rCBV (Figure 3B). Notably, our main intention is to obtain the DSC-MRI-derived rCBV without GBCA injection. The $T_2w$ images are generally obtained from plain scans, thus their participation does not change the nature of the work. Le Bihan et al. have proposed that the IVIM effect affects the diffusion MRI signals at low diffusion weighting ($b < 200$ s/mm$^2$) [28], and Pang et al. have concluded that the tumor perfusion fraction (IVIM-$f$) could be indistinguishable from normal with the inclusion of higher $b$-value DWI images where the contribution due to non-Gaussian diffusion was appreciable [29]. However, it can be seen from the results in Figure 3 that the diffusion MRI signals at high diffusion weighting still improves the SSIM of synthesized rCBV. The potential explanation might be that although the signal intensity of DWI images with high *b* value attenuates much, they are more sensitive to proton motion, carry more proton flow information, and provide more information for synthesizing rCBV. Although, according to Eq. (1), the rCBV map could be synthesized with only *f* and $f_W$ images, results in Figure 3 reveal that the neural network prediction with only IVIM-*f* and T$_2$w as input is less than satisfactory. It may be explained with the following reasons: First, the utilization of IVIM-*f* is equivalent to the information compression of multi-*b*-value DWI images, which inevitably reduces the effective features, i.e., the information of IVIM-*D* and IVIM-*D*$^*$, since researchers have concluded that IVIM parameters and rCBV have moderate correlations [26]; Second, there may be certain errors in the calculated IVIM-*f* because of the great difficulty of IVIM-*f* calculation due to the low signal-to-noise ratio of IVIM imaging.

Although not displayed in the present work, we have experimented with various neural network frameworks to determine the final selection for the proposed framework. Pix2pix, cycleGAN and transformer have all been applied to the current framework but achieved worse performance with higher computational consumption. The low suitability between the complexity of the conversion task and neural network architecture, resulting in possible overfitting, could be a potential explanation. The networks mentioned above are frequently used to convert completely unrelated images, which differs from the conversion of multi-*b*-value DWI to rCBV. Moreover, the small amount of training data is another limitation for larger-scale networks like the transformer, which greatly demands training data quantity. Actually, Asaduddin et al. have recently fulfilled the DSC-MRI derived parametric maps generation from dynamic MR angiography with U-Net and its fine-tuned version [4]. That certifies, to some extent, the feature extraction ability of U-Net and its variant is sufficient for some modality transformation problems with low difficulty. Furthermore, Li et al. have pointed out that the benefits of GANs for image synthesis in biomedical imaging must be carefully evaluated, especially for quantitative imaging [23]. In their work, U-Net synthesized T$_1$w images outperform the ones synthesized from GAN in better-improving brain segmentation accuracy. Apart from employing theoretically more powerful networks, minor network structure modifications, including enlarging the encoder/decoder size, adding the residual block, and adding the attention block, are also done before the final model is determined. While according to statistical analysis, these modifications slightly influence the concordance or SSIM between the synthetic and real rCBV maps. Actually, the introduced perceptual loss contributes a lot in the proposed framework, it emphasizes the image details in U-Net predicted synthetic rCBV maps and largely reduces the image over-smoothing.

Compared with the other two rCBV synthetic methods [4; 22], the structural similarity of the rCBV maps obtained by our proposed method relative to real ones (maximum SSIM = 0.83) is lower than the rCBV synthesized from MRA (SSIM = 0.89 ± 0.08). This may be due to the fact that the local magnetic field strength change caused by contrast agent injection is more sensitive to the proton flow than the diffusion-sensitive gradient. Richer proton flow information might be obtained by enhancing the signal-to-noise ratio of images acquired at high diffusion-sensitive gradients. And the T/WM ratio linear regression analysis result of our work (IDH wildtype: slope = 0.74; $R^2$ = 0.74; Pearson $P$ = 0.75; IDH mutation: slope = 0.77; $R^2$ = 0.79; Pearson $P$ = 0.89) is comparable to the DCE-MRI synthesized rCBV (slope = 0.72; $R^2$ = 0.57; Pearson $P$ = 0.86).

Synthetic rCBV has several advantages. First, it enables rCBV map obtaining without contrast agent injection and has similar performance to DSC-MRI derived rCBV in the identification of IDH mutation and wildtype glioma, avoiding potential kidney impairment and allergic syndrome. Second, synthetic rCBV reduces parameter estimation time (from conventional deconvolution to deep learning algorithm) and improves visual quality of rCBV map. Third, synthetic rCBV is available for motion artifact reduction, as shown in Figure 4B (the eighth column). The potential reason is that network training labels are almost artifact-free, and the predictions would follow the learned prior knowledge. Fourth, it shows prominent IDH mutation status identification capacity, assisting the glioma therapy formulation and prognosis.

There are several limitations in the present study. First, the IDH mutation cases are limited, which hampers the proposed framework's performance and further evaluation of the diagnostic value of the framework. We believe that with more IDH mutation cases, in other words, more prior information, for neural network training, the IDH mutation data prediction would be more precise and better IDH status identification would be implemented. Second, our validation is based on a single center, and external validations should be conducted on multicenter data to verify the generalization ability of the proposed framework. In the future, more IDH mutation data of multi-center should be collected to achieve better performance and stronger robustness of the proposed framework. Furthermore, as well-established, CBV is also a significant indicator for stroke diagnosis and offers vessel architecture information for planning reperfusion treatment in stroke with vessel occlusion [30; 31]. Hence, work could be done on stroke cases to further confirm the clinical value of the proposed framework.

In conclusion, it is feasible to acquire DSC-MRI-derived rCBV without injection of GBCA, and the synthetic rCBV has comparable glioma IDH mutation status identification ability as real rCBV, allowing the perfusion information acquisition for kidney impaired or pregnant patients and benefits the treatment follow up evaluation.

## Acknowledgments

This work was supported by the National Natural Science Foundation of China under grant numbers 82071913 and 22161142024, in part by the Science and Technology Project of Fujian Province of China under Grant 2021Y9154.

# Supplemental materials

**Real rCBV**

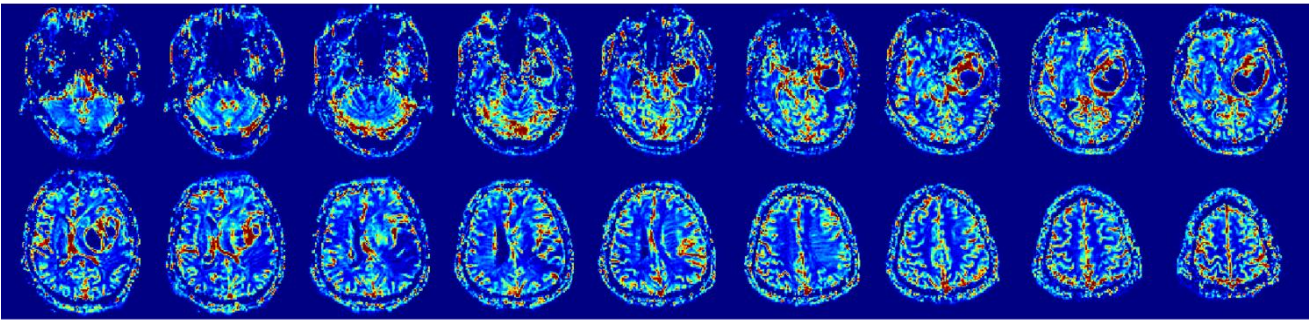

**Synthetic rCBV**

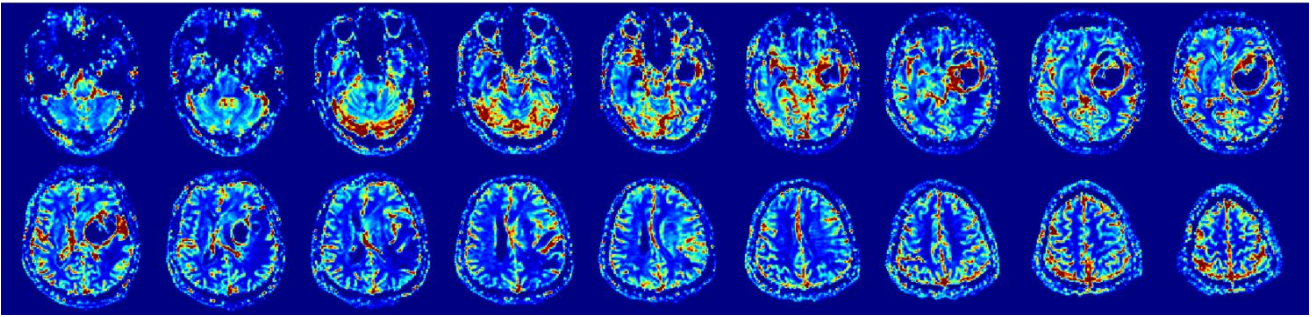

Figure S1. The real and synthetic rCBV maps of the whole brain (18 slices) of IDH wildtype patient #2.

**Real rCBV**

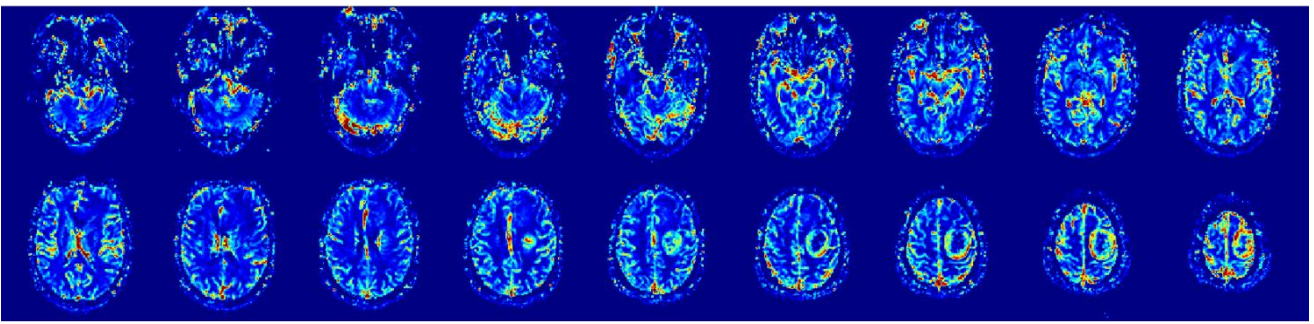

**Synthetic rCBV**

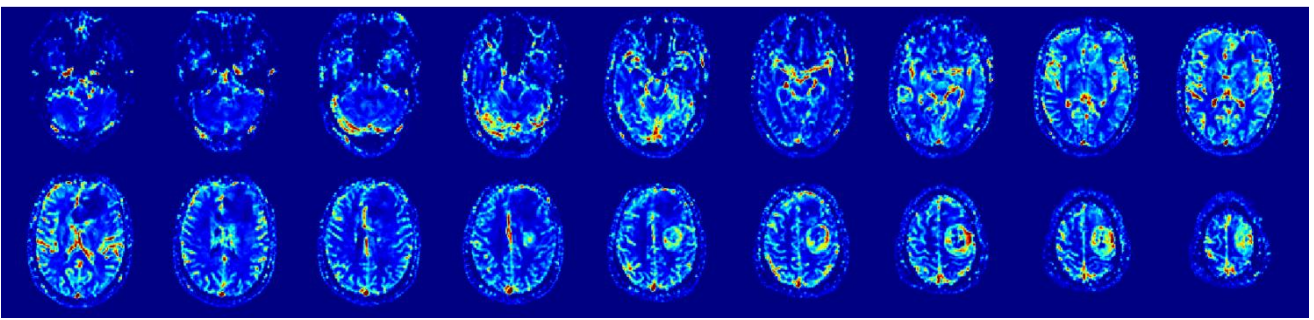

Figure S2. The real and synthetic rCBV maps of the whole brain (18 slices) of IDH mutation patient #2.